\documentclass[12pt]{article}
\usepackage{geometry}
\usepackage{setspace}
\usepackage[numbers]{natbib}
\usepackage{doc}
\usepackage{url}
\usepackage{graphicx}
\usepackage{epstopdf}
\usepackage{hyperref}
\usepackage{amsthm}
\usepackage{mathtools}
\usepackage[font=small,labelfont=bf]{caption}
\providecommand{\keywords}[1]{\textbf{\textit{Keywords---}} #1}
\title{An APX for the Maximum-Profit Routing Problem with Variable Supply 
}
\author{Bogdan Armaselu
\footnote{Email: barmaselub@gmail.com}
}
\date{}

\begin{document}

\maketitle
\begin{center}

\end{center}

\abstract{
In this paper, we study the Maximum-Profit Routing Problem with Variable Supply (MPRP-VS). 
This is a more general version of the Maximum-Profit Public Transportation Route Planning Problem, or simply Maximum-Profit Routing Problem (MPRP), introduced in \cite{Armaselu-PETRA}. 
In this new version, the quantity $q_i(t)$ supplied at site $i$ is linearly increasing in time $t$, as opposed to \cite{Armaselu-PETRA}, where the quantity is constant in time. 
Our main result is a $5.5 \log{T}  (1 + \epsilon) (1 + \frac{1}{1 + \sqrt{m}})^2$ approximation algorithm, where $T$ is the latest time window and $m$ is the number of vehicles used.
In addition, we improve upon the MPRP algorithm in \cite{Armaselu-PETRA} under certain conditions.
 }


\section{Introduction}
\label{s:intro}

In this paper, we consider the Maximum-Profit Routing Problem with Variable Supply (MPRP-VS), stated as follows.
Consider $n$ sites $S = \{s_1, \dots, s_n\}$ in the euclidean plane, 
each site $s_i$ being specified by its coordinates $x_i, y_i$, the production rate $\rho_i \in Z$ of a certain good, and an interval $[e_i, l_i]$ (in hours) called \textit{time window}. 
The values $e_i, l_i$ are assumed to be integers in the interval $[0, T]$, where $T$ is a given constant, and supply loading is assumed to be instantaneous (that is, no service time needed).
For every site $s_i$, the quantity $q_i(t)$ of the good available at time $t$ at site $i$, is assumed to be 0 outside the time window, and piecewise linearly increasing at the rate $\rho_i$ within the time window.
The distance $d_{i,j}$ between sites $s_i$ and $s_j$ is the euclidean distance.
We are also given a fleet of $m$ vehicles $V = \{v_1, \dots, v_m\}$, each having the same capacity $Q$.
All vehicles are assumed to travel at unit speed and have unit fuel consumption per distance travelled.
Only one vehicle may visit any given site and all vehicles must start and end their tour at the same given depot $D(x_D, y_D)$.
The goal is to compute, for every vehicle $v_k$, a route $r_k$ that collects a quantity $q^k_i$ of the good such that the total profit of all routes is maximized, 
where the profit of a route $r_k$ is the total quantity collected $\sum_{s_i \in r_k}{q^k_i}$ (called \textit{reward}) minus the total distance travelled via $r_k$ (called \textit{costs}).

MPRP-VS is a generalization of Maximum-Profit Pick-up Problem with Time Windows and Capacity Constraint (MPRP), introduced in \cite{Armaselu-PETRA} (see also \cite{Armaselu-arXiv}.
MPRP is similar to the problem considered in this paper, the difference being that the quantity of the good supplied at each site is constant in time.
In \cite{Armaselu-PETRA}, two polynomial-time approximation algorithms with constant approximation factors are given for MPRP.
They both run in $O(n^{11})$ time, assuming $q_i(l_i) \leq \alpha q_j(l_j), \forall{i, j = 1, \dots, n}$, for some constant $\alpha > 1$.
The best performance ratio, $15 \log T$, among the two algorithms, is achieved by an approach involving well-separated pair decompositions.

MPRP-VS is a variant of the Euclidean Travelling Salesman Problem (TSP), which is known to be strongly NP-hard \cite{Dantzig, Miller}. 
Hence, MPRP-VS is also a strongly NP-hard problem, which means there is no algorithm with a running time polynomial in the value of the input unless P = NP.

Applications of the results presented in this paper include public transportation in various settings, such as cities \cite{Armaselu-PETRA}, inter-city railway transportation, domestic or international flight, etc.
Another important application is in for-profit waste pick-up, in which the reward is proportional to the amount of waste collected.

\subsection{Related work}
\label{ss:related-work}

Various cases and generalizations of TSP have been studied, and various approximation algorithms and inapproximability results have been given.
It is known that general TSP cannot be approximated within any constant factor \cite{Sahni}.
However, constant factor approximations have been proposed for some restricted versions.
For metric TSP, the best-known result is Christofides's algorithm \cite{Christofides}, which runs in $O(n^3)$ time for an instance of $n$ sites, and has an approximation factor of $3/2$.
For the euclidean case, Arora et. al \cite{Arora} presented an $O(n (\log n)^{O(1/\epsilon)})$-time randomized algorithm for the planar case, with a $(1 + \epsilon)$ approximation ratio. 
They also show how to extend their algorithm for the $d$-dimensional case, in which case their running time increases to $O(n (\log n)^{O((\sqrt{d} / \epsilon)^{d-1})})$.

Fisher et. al studied the Vehicle Routing Problem with Capacity Constraint (VRP) \cite{Fisher94}, 
a generalization of TSP in which there are $m$ vehicles of equal capacity $Q$, there is only one depot for all vehicles, and each customer $i$ has a demand $q_i$ of a specified good. 
They gave a near-optimal iterative algorithm for the problem, using an iterative lagrangian relaxation of the constraints.
Later, study the VRP with Time Windows (VRPTW) \cite{Fisher95}, which is a generalization of VRP with time windows $[e_i, l_i]$ and different capacity constraints $Q_j$ for the vehicles. 
They adapt their algorithm in \cite{Fisher94} to this version, as well as give an additional linear programming approach.

Golden et. al introduced the Capacitated Arc Routing Problem (CARP) \cite{Golden}, 
in which every edge has a demand and a cost, every vehicle has the same capacity, and the goal is to minimize total cost while meeting all demands.
They prove that CARP can be approximated within a constant factor when the triangle inequality is satisfied.
Later, van Bevern extended the result to general undirected graphs \cite{Bevern}.

Wang et. al \cite{Wang} considered a somewhat similar problem, namely the Multi-Depot Vehicle Routing Problem with Time Windows and Multi-type Vehicle Number Limits \cite{Wang}. 
The main difference to our problem is that the goal is to minimize the number of vehicles used (if feasible) or maximize the number of visited customers (if infeasible). 
Although they solve the problem using a genetic algorithm approach, their algorithm is iterative and has no performance guarantee with respect to the approximation factor.

The Deadline-TSP problem was studied by Bansal et. al \cite{Bansal}. 
In Deadline-TSP, we are given $n$ sites, where each site $i$ has a deadline $D_i$ and a reward $r_i$, and we want to find a tour that maximizes the total reward of sites that can be visited. 
They propose an algorithm with an approximation factor of $O(\log n)$. 
They also show how to extend it to the more general TSP with Time Windows (TSPTW), in which the sites also have release times $R_i$. 
For TSPTW, they give an algorithm with an $O(\log^2 n)$ approximation. 
They also give an $O(\log(1/\epsilon))$-approximate algorithm for the $\epsilon$-relaxed version of the problem, in which we are allowed to extend the deadlines by a factor of $1 + \epsilon$.

\subsection{Challenges}
\label{ss:challenges}

The problem we study involves major challenges.

First, the reward maximization part of our solution relies on a bi-criteria solver for the Travelling Salesman Problem with time windows (TSPTW), 
which is known to be hard to approximate within a constant factor $c$ for every $c > 1$ in the general case \cite{Bockenhauer}.
In the case where time windows are integer, Bansal et. al \cite{Bansal} show how to solve TSPTW within an $O(\log T)$ approximation factor, where $T$ is the latest deadline.
However, even for constant $T$, the constants involved in their approximation factor are still very large.

Second, it is not sufficient to maximize rewards, since the highest reward tour may have high travelling costs, which may render the optimal-reward tour not profit-optimal.
Moreover, if a vehicle takes too long to complete its tour, it may miss time windows of sites it may otherwise be able to visit under the reward-optimal tour.

Third, a reward-optimal tour that takes into account only the maximum or average quantity supplied at each site may not even be reward-optimal for the variable-supply version,
as a site visited later gives a better reward, but visited too late gives no reward.

\subsection{Our contributions}
\label{ss:structure}

We give a constant-factor approximation algorithm for MPRP-VS, which uses clever reductions to some known problems.
Specifically, our algorithm runs in $O((\frac{n}{\epsilon^2})^{11})$ time and has an approximation ratio of $5.5 \log{T} (1 + \epsilon) (1 + \frac{1}{1 + \sqrt{m}})^2$.
In addition, our algorithm can be adapted to the MPRP problem to improve upon the algorithm in \cite{Armaselu-PETRA}, under certain assumptions about the quantities supplied.

The rest of the paper is structured as follows.
In Section \ref{s:prelim}, we formulate the problem in formal, mathematical ways, make some preliminary observations, and list some reasonable assumptions about the given data.
Then, in Section \ref{s:restricted}, we solve the problem on a restricted one-vehicle case, show how to reduce to the problem in \cite{Armaselu-arXiv}, and analyze the algorithm.
After that, in section \ref{s:general}, we generalize to multiple vehicles and provide the APX for the general case.
Finally, in Section \ref{s:conclusion}, we conclude and list some future directions.

\section{Preliminaries}
\label{s:prelim}

MPRP-VS can be mathematically formulated as:

Maximize $\sum_{k = 1, \dots, m}{\sum_{s_i \in r_k}{q^k_i} - \sum{(s_i, s_j) \in r_k}{d_{i,j}} - d_{0,s^k_s} - d_{s^k_e,0}}$ (1)

subject to

$s^k_s, s^k_e \in r_k, \forall{k = 1, \dots, m}$ (2),

$\sum_{s_i \in r_k}{q^k_i} \leq Q, \forall{k = 1, \dots, m}$ (3),

$e_i \leq t^k_i \leq l_i, \forall{k = 1, \dots, m; s_i \in r_k}$ (4),

$\forall{k = 1, \dots, m; (s_i, s_j) \in r_k}, t^k_j - t^k_i \leq d_{i,j}$ (5),

$q^k_i = \rho_i (t^k_i - e_i), \forall{k = 1, \dots, m; s_i \in r_k}$ (6), and

$\forall{k, l = 1, \dots, m: k \neq l; 1 \leq i \leq n}, s_i \in r_k \implies s_i \notin r_l$ (7),

where $t^k_i$ is the timestamp when vehicle $v_k$ visits site $s_i$.

Here
(1) is the profit maximization objective function,
(2) is the first and last site visitation constraint,
(3) is the capacity constraint,
(4) is the time window constraint,
(5) is the travelling constraint,
(6) is the production constraint,
and (7) is the per-site vehicle uniqueness constraint.
Note that the depot $D$ is treated as the 0-th site.

We call the above \textit{Formulation 1}.

Alternatively, one can formulate the problem as follows.

Max. $\sum_{0 \leq i \leq n} \sum_{1 \leq k \leq m} \sum_{0 \leq t \leq T} (\sum_{1 \leq j \leq n} x_{i,j,k,t} q_j(t) - \sum_{0 \leq j \leq n} x_{i,j,k,t} d_{i,j})$ (1)

subject to

$\sum_{0 \leq i \leq n} \sum_{1 \leq j \leq n} \sum_{0 \leq t \leq T} x_{i,j,k,t} q_j(t) \leq Q, \forall{k = 1, \dots, m}$ (2),

$q_j(t) = 0, \forall{1 \leq j \leq n, 0 \leq t \leq T: t \notin [e_j, l_j]}$ (3),

$q_j(t) = \rho_i (t - e_j), \forall{1 \leq j \leq n, t \in [e_j, l_j]}$ (4),

$\sum_{0 \leq i \leq n} \sum_{1 \leq k \leq m} \sum_{0 \leq t \leq T} x_{i,j,k,t} = 1,  \forall{1 \leq j \leq n} $ (5),

$\sum_{0 \leq i \leq n} \sum_{0 \leq t \leq T} x_{i,j,k,t} = \sum_{0 \leq i \leq n} \sum_{0 \leq t \leq T} x_{j,i,k,t}, \forall{1 \leq k \leq m, 1 \leq j \leq n} $ (6),

$\sum_{1 \leq j \leq n} \sum_{1 \leq k \leq m} \sum_{0 \leq t \leq T} x_{0,j,k,t} = m$ (7.1),

$\sum_{1 \leq i \leq n} \sum_{1 \leq k \leq m} \sum_{0 \leq t \leq T} x_{i,0,k,t} = m$ (7.2),

where $x_{i,j,k,t} = 1$ if vehicle $v_k$ arrives from $s_i$ at $s_j$ at time $t$.

Here
(1) is the profit maximization objective function,
(2) is the capacity constraint,
(3) is the time window constraint,
(4) is the production constraint,
(5) is the per-site vehicle uniqueness constraint,
(6) is the tour connectivity constraint,
and (7.1), (7.2) are the constraints enforcing starting and ending at the depot.

We call this alternative \textit{Formulation 2}.

We assume that all sites start "empty", that is, $q_i(e_i) = 0, \forall{1 \leq i \leq n}$.
Further, we suppose the discrepancies in the quantities supplied are small,
specifically $q_i(l_i) \leq \alpha q_j(l_j), \forall{i, j = 1, \dots, n}$, for some constant $\alpha > 1$, and $Q, q_i(l_i) \in Z, \forall{1 \leq i \leq n}$.

Since the quantities supplied increase linearly in time, we have $q_i(t) = \rho_i (t - e_i) = q_i(l_i) \frac{t - e_i}{l_i - e_i}$.

Note that, if $Q \geq \sum_{i=1}^{n} q_i(l_i)$, then the capacity constraint is always satisfied.
Thus, from now on, we also assume $Q < \sum_{i=1}^{n} q_i(l_i)$.

\section{One-vehicle case}
\label{s:restricted}

We first consider a restricted version of the MPRP-VS problem, say MPRP-VS-1, in which only one vehicle is available and show how to reduce it to the MPRP problem with one vehicle (MPRP-1).

\textbf{Lemma 1}. MPRP-VS-1 problem can be reduced to MPRP-1 within $O(\frac{n}{\epsilon})$ time and a ratio of $(1+\epsilon)$, for any $\epsilon > 0$. 
That is, the vehicle assigned for an instance $I$ of MPRP-VS-1 collects a reward $r^I \leq (1+\epsilon) r^J$, 
where $r^J$ is the quantity collected by the vehicle assigned for $J$ and $J$ is the corresponding instance of MPRP-1.

\textbf{Proof}. Let $I$ be an instance of the restricted MPRP-VS-1 problem. 
We construct an instance $J$ of MPRP-1. 
Let $\epsilon > 0$.
For every site $i$ on instance $I$, we consider the $N$ intervals $I_{i,1}, \dots, I_{i,N}$, with 

$I_{i,1} = [e_i, e_i + \frac{l_i - e_i}{(1+\epsilon)^{N-1}}], I_{i,\tau} = [e_i + \frac{l_i - e_i}{(1+\epsilon)^{N-\tau+1}}, e_i + \frac{l_i - e_i}{(1+\epsilon)^{N-\tau}}], \forall \tau = 2 ... N$ .

Here $N$ is the smallest integer such that $(1+\epsilon)^{N-1} \geq 1 / \alpha$, that is, $N = 1 + \frac{\ln (1 / \alpha)}{\epsilon}$
(see Figure 2 for an illustration). 
Note that, the length of the interval containing timestamp $t$ is a non-decreasing function of $q_i(t)$. 
Indeed, $|I_{i, 1}| = \frac{l_i - e_i}{(1+\epsilon)^{N-1}}$, and $|I_{i, \tau}| = \frac{\epsilon (l_i - e_i)}{(1+\epsilon)^{N-\tau+1}}, \forall \tau > 1$, which is increasing in $\tau$.
For each interval $I_{i,\tau}$, we add a site labeled $(i, \tau)$ to $J$, with the same coordinates as site $i$ and with fixed quantity $q(i, \tau) = q_i(l_i) \frac{\tau - 0.5}{N - 1}$ and time window $J_{i,\tau} = I_{i,\tau}$. 
For every two nodes $(i, \tau), (j, \tau')$ in $J$, the point-to-point distance $d_J((i, \tau), (j, \tau'))$ is defined as follows. 

$d_J((i, \tau), (j, \tau')) = 
\begin{cases}
    d_I(i, j),& \text{if } i \neq j \\
    0,              & \text{otherwise}
\end{cases}
$

\begin{figure}
\begin{center}
	\includegraphics[scale=0.2]{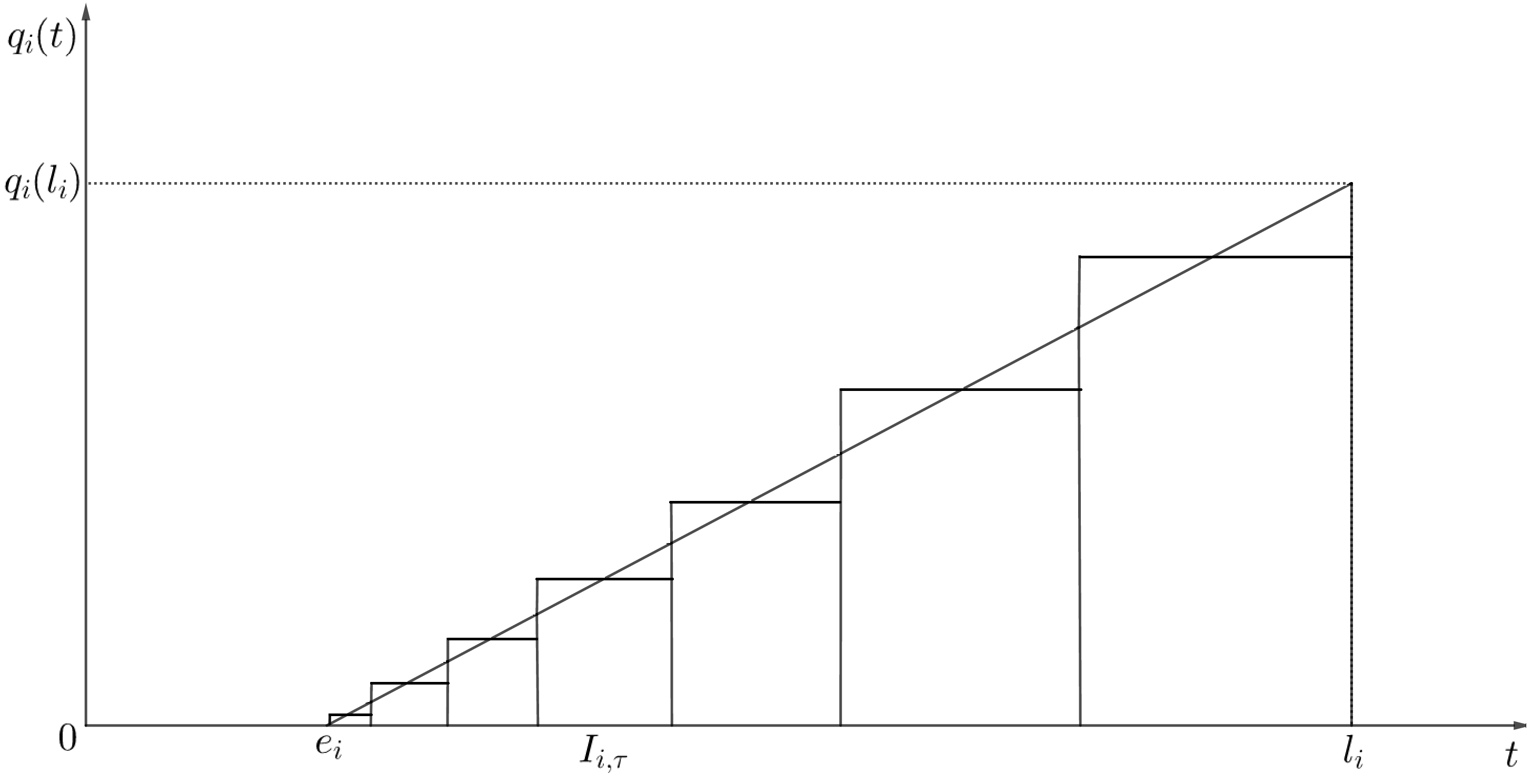}
	\caption{The intervals $I_{i, \tau}$ for $\alpha = e^{-3}, \epsilon = 0.5, \tau = 1 \dots 7$}
\end{center}
\end{figure}

After constructing $J$, we approximately solve the optimization Subset-Sum problem, in which the goal is to find a set of numbers that add up to as close as possible to $Q$ without exceeding $Q$.
The input set of numbers for Subset-Sum will be $S = \{ q(i, \tau) | i = 1 \dots n, \tau = 1 \dots N \}$.
Denote by $S^J$ the set of numbers output by the Subset-Sum solver. 

Using the reward-maximization algorithm in \cite{Bansal}, we compute a routing $R^J$ on $J$. 
Then, for every $i = 1 \dots n$, we take the latest interval $\tau_i$ for which $(i, \tau_i) \in R^J$, and remove all nodes $(i, \tau) \in R^J$ such that $\tau \neq \tau_i$.,
to ensure that every site is visited only once.
The resulting route is $R^I = \{ i | \exists \tau: (i, \tau) \in R^J \}$.

By construction, the list of input nodes in $R^I$ is the same as the one in $R^J$. 
Moreover, for each site $i$, the time interval $I_{i, \tau_i}$ in which $R^I$ visits $i$ is the same as the time window $J_{i, \tau_i}$. 
Also, we solve the Subset-Sum problem optimally in $O^*(\min{Q \sqrt{n}, Q^{4/3}, \sum_i{q(i, \tau)}})$, using the method in \cite{Koiliaris}, where $O^*$ ignores a polylogarithmic factor.
Since $\sum_i{q(i, \tau)} = O(\frac{\rho}{\epsilon}  n T)$, the running time becomes $O^*(\frac{\rho}{\epsilon} n T) = O^*(\frac{n}{\epsilon})$ time.
The quantity $\rho^I$ collected by $R^I$ is at least $\rho^J \cdot \beta$, with $\beta = sup_{i,\tau}{\frac{q_{i, max}(\tau)}{q_{i, min}(\tau)}}$, 
where $q_{i, max}(\tau)$ and $q_{i, min}(\tau)$ are the maximum (respectively, minimum) quantity collectible in interval $\tau$ of site $i, \forall i, \tau$. 
It is easy to check that $\beta = 1 + \epsilon$ by construction. 
Hence, the result follows. \qed

\section{General case}
\label{s:general}

Now we show how to extend the reduction to solve the general MPRP-VS problem (for multiple vehicles). 

Given an instance $I*$ of MPRP-VS, we do the following.

1. Decompose $S$ into a set of subsets $S_k$ for each vehicle $v_k$.
For each subset $S_k$, create an instance $I_k$ of MPRP-VS-1 with the set $S_k$ of sites, the same capacity constraint $Q$ and the same time windows.

2. For each instance $I_k$, construct an instance $J_k$ of MPRP-1 from $I_k$, using the reduction in Section \ref{s:restricted}.

3. Solve each instance $J_k$ and denote by $r_k$ the resulting route.

4. For each site $s_i$, if two or more vehicles visit $s_i$, retain only the vehicle $v_k$ that would collect the largest $q_i^k$.

5. For each site $s_i$ and vehicle $v_k$, reassign $s_i$ to $r_k$ if and only if the profit would increase. 

6. Set $S = S - \cup_k{S_k}$

7. Repeat steps 1-6 until either there are no sites left unassigned or no more sites are assigned during an iteration.

Now we describe Steps 1 and 3 in detail.

\subsection{Decomposing $S$ into subsets}
\label{ss:decomposition}

We first construct a well-separated pair decomposition (WSPD) $W$ of the sites, using the separation factor $s = \sqrt{m}$.
Note that $W$ is a binary tree-like hierarchical clustering of the sites such that, for any pair $(s_i, s_j)$ of sites, exactly one pair $(A_k, B_k)$ of subsets in $W$ $s$-separates them.
That is, either $s_i \in A_k, s_j \in B_k$ or viceversa.
There are $O(s^2 n)$ subsets in $W$, and each subset $A_i$ or $B_i$ is further decomposed into $O(s^2)$ pairs of subsets $(A_{i+1}, B_{i+1}), \dots, (A_{i+s^2}, B_{i+s^2})$.
Since this is a hierarchical pair decomposition, there are $1 + O(\frac{\log n}{\log s})$ levels in the hierarchy, and the union of the subsets on any level in the whole input set $S$.
Assume wlog that the pairs of subsets are ordered by the total number of sites contained in them.

After that, for every vehicle $v_k$, we construct an instance $I_k$ of MPRP-VS-1 where the set of sites is $S_k$ assigned to $v_k$ as follows.

For vehicles $v_1, v_2$, assign subsets $S_1 = A_1, S_2 = B_1$.
For every $k > 1$, for vehicles $v_{2k-1}, v_{2k}$, assign subsets $S_{2k-1} = A_{k}, S_{2k} = B_{k}$, 
and remove all sites in $S_{2k-1}, S_{2k}$ from $S_1, \dots, S_{2k-2}$.

It is easy to check that no subsets are overlapping.

Moreover, since the subgraph induced by the subsets is an $s$-spanner, 
it follows that any minimum-length routing algorithm on the induced subgraph outputs a total length of at most $L = L^* \cdot (1 + \frac{1}{1 + s})$,
where $L^*$ is the cost output by the algorithm when run on the input set of sites.

This gives us the following result.

\textbf{Lemma 2}.
The total length of a minimum-length routing $r_k$ for each subset $S_k$ is at most $L = L^* \cdot (1 + \frac{1}{1 + \sqrt{m}})$,
where $L^*$ is the total length of a minimum-length routing on $S$.

\subsection{Solving an instance $J_k$}
\label{ss:solving-mppr-1}

First, we compute a route $r_k'$ that approximately maximizes the quantity collected from $S_k$, using the algorithm in \cite{Bansal}.
After that, we traverse $r_k'$ and, for every two consecutive arcs $(s_i, s_j), (s, j, s_l)$, replace them by a single arc $(s_i, s_l)$ if and only if it meets and time windows and increases the profit.
Denote the resulting route by $r_k$.

\textbf{Lemma 3}. 
$r_k$ has a profit $P_k \geq  \frac{P_k^*}{8 \ln{2} \log{T}}$, 
where $P_k^*$ is the maximum profit obtainable by a tour visiting sites in $S_k$ with capacity constraint $Q$.

\textbf{Proof}.
With the algorithm in \cite{Bansal}, $r_k'$ picks up a total quantity of at least $\frac{Q_k^*}{8 \ln{2}\log{T}}$,
where $Q_k^*$ is the total quantity collected by an optimal algorithm.

%After that, we choose $p$ arcs from $r_k$ and re-order them in such a way as to maximize total profit.

%Since there $O(n^p)$ arrangements of arcs that we need to consider, we can stay within the $O(n^{11})$ running time bound in \cite{Armaselu-PETRA} by choosing $p = 11$.

%Note that this arc rearrangement process also improves upon the profit maximization approach in \cite{Armaselu-PETRA} in the general case when there is no assumption about the distance travelled-to-revenue ratio.
%That is, regardless of the distance matrix $d$, we achieve improved bounds on the profit using this arc rearrangement.

After arc replacements on $r_k'$, the total profit of the resulting route $r_k$ becomes 

$\sum_{s_i \in r_k}{q_i} - \sum_{(s_i, s_j) \in r_k}{d_{ij}} \geq \sum_{s_i \in r_k'}{q_i} - \sum_{(s_i, s_j) \in r_k'}{d_{ij}} \geq \frac{Q^*_k}{8 \ln{2}\log{T}} - \sum_{(s_i, s_j) \in r_k'}{d_{ij}}
\geq \frac{Q_k^*}{8 \ln{2}\log{T}} - L_k^* \geq \frac{Q_k^* - L_k^*}{8 \ln{2}\log{T}} \geq \frac{P_k^*}{8 \ln{2}\log{T}}$,

where $L_k^*$ is the total length of the shortest tour visiting all sites in $S_k$ with capacity constraint $Q$.

%That is, the approximation factor on the profit is $8 \ln(2)$.

%We use the Well Separated Pair Decomposition-based (WSPD) algorithm in \cite{Armaselu-PETRA} to solve $J^*$. 
%After the WSPD-based algorithm computes a routing $R^*$ on $J^*$, 
%for every $i = 1 \dots n$, we take the latest interval $\tau_i$ for which $(i, \tau_i) \in R^*$ and discard all nodes $(i, \tau) \in R^*: \tau \neq \tau_i$. 
%Finally, we discard all vehicles with empty routes.

\subsection{Combining routes and reassigning sites}
\label{ss:combining}

We analyze 
Let $S_t$ be $S$ at the beginning of the $t$-th iteration.

\textbf{Lemma 4}. 
During some iteration $t$, after Step 5 of the algorithm, the total quantity collected from $S_t$ is at least $\frac{Q_t^*}{(1 + \frac{1}{1 + \sqrt{m}})(1 + \epsilon)}$, 
where $Q_t^*$ is the optimal quantity collectible from $S_t$.

\textbf{Proof}.
Suppose on the contrary that, in any iteration $t$, the total quantity $Q_t$ collected after Step 5 is less than 
$\frac{Q_t^*}{(1 + \frac{1}{1 + \sqrt{m}})(1 + \epsilon)}$.
Then there exists a site $s_i$ from which a quantity less than $\frac{q_i^*}{(1 + \frac{1}{1 + \sqrt{m}})(1 + \epsilon)}$ is collected, 
where $q_i^*$ is the quantity collected from $i$ by the optimal MPRP-1 algorithm.
By Lemma 1, it follows that the optimal MPRP-VS-1 algorithm collects a quantity less than $\frac{q_i^*}{1 + \frac{1}{1 + \sqrt{m}}}$.
Since the travel costs on $S_t$ are at most $1 + \frac{1}{1 + \sqrt{m}}$ times the optimal value,
it follows that the algorithm misses a site $s_i$ satisfying $q_i^* > \frac{Q_t^*}{n}$ due to missing its time window.
However, by Step 7, if a site is missed during an iteration, it will eventually be visited during a further iteration $t'$.
\qed

\subsection{Analysis}
\label{ss:analysis}

We now analyze the performance of the algorithm. 

By Lemma 2, the total distance travelled by all $v_k$ on all $S_k$ is at most $1 + \frac{1}{1 + \sqrt{m}}$ times higher than the optimal distance.
By Lemma 1, for each $v_k$, the quantity collected from all sites in $S_k$ is at least $\frac{1}{1 + \epsilon}$ of the quanity collected by an MPRP-1 solver on $S_k$.
Since $r_k$ is the output of an MPRP-1 solver on $S_k$ and, by Lemma 3, the profit collected by $r_k$ from $S_k$ has a profit at least $\frac{1}{8 \ln{2}\log{T}}$ the optimal profit,
it follows that the profit for each $v_k$ is at least $\frac{1}{1 + \epsilon} \cdot \frac{1}{8 \ln{2}\log{T}}$ times the optimal for an MPRP-VS-1 solver on $S_k$.
Now the profit of the algorithm on the original problem instance $I$ would be $\frac{1}{1 + \epsilon} \cdot \frac{1}{8 \ln{2}\log{T}}$ times the optimal in absence of time winodws.
The extra tour length factor of $1 + \frac{1}{1 + \sqrt{m}}$ may cause more than $1 + \frac{1}{1 + \sqrt{m}}$ as much revenue to be lost due to missing time windows.
However, by Lemma 4, we collect a total quantity at least $\frac{Q^*}{(1 + \frac{1}{1 + \sqrt{m}})(1 + \epsilon)}$ while increasing travel costs at most $1 + \frac{1}{1 + \sqrt{m}}$ times,
where $Q^*$ is the optimal total quantity collectible.

That is, the total profit for the algorithm becomes at least 

$P = P^* \frac{1}{(1 + \epsilon)^2} \cdot \frac{1}{8 \ln{2}\log{T}} \cdot \frac{1}{(1 + \frac{1}{1 + \sqrt{m}})^2} = \frac{P^*}{8 \ln{2}\log{T} (1 + \epsilon)^2 (1 + \frac{1}{1 + \sqrt{m}})^2}$,

where $P^*$ is the optimal profit.

Letting $\delta = 2\epsilon + \epsilon^2$, we get 

$P = \frac{P^*}{8 \ln{2}\log{T} (1 + \delta) (1 + \frac{1}{1 + \sqrt{m}})^2}$.

Additionally, note that Step 2 produces a routing on a total of $O(\frac{n}{\epsilon})$ sites.
Since Step 3 takes $O(N^{11})$ time when run on a et of $N$ sites \cite{Armaselu-PETRA}, it runs in $O((\frac{n}{\epsilon})^{11})$ time in our case.

Since $\epsilon = -1 + \sqrt{1 + \delta^2} = \frac{\delta^2}{1 + sqrt{1 + \delta^2}} = \Theta(\delta^2)$, we get a ruinning time of $O((\frac{n}{\delta^2})^{11})$.

Re-assigning $\delta$ as $\epsilon$, we can state the following result.

\textbf{Theorem 5}. For any parameter $\epsilon$, MPRP-VS can be solved within $O((\frac{n}{\epsilon^2})^{11})$ time and has an approximation ratio of 

$8 \ln{2} \log{T} (1 + \epsilon) (1 + \frac{1}{1 + \sqrt{m}})^2 \simeq 5.5 \log{T} (1 + \epsilon) (1 + \frac{1}{1 + \sqrt{m}})^2$.

%By cleverly choosing the parameter $\epsilon$, we can get a suitable running time / approximation factor trade-off. By choosing $s = \sqrt{n}, \epsilon = \frac{1}{\ln{1 / \alpha}}$, we get a running time of $O(n^{11+p})$ and an approximation ratio of $11 (1 + \frac{2 \epsilon}{1 - \alpha}) \log T$ since $\beta \leq \frac{2}{1 - \alpha}$.

Notice that the algorithm presented in this paper can also be used to solve MPRP.
When the quantities supplied at each site are constant in time, step 2 of the algorithm can be skipped, and step 3 can be run on $I_k$ instead of $J_k$.
Since the extra $\frac{1}{(1 + \epsilon)^2}$ factor in the approximation ratio and the extra $O(\frac{1}{\epsilon^{22}}$ is due to the reduction used in step 2, we obtain the following.

\textbf{Corollary 6}.
MPRP can be solved in $O(n^{11})$ time and has an approximation factor of $\simeq 5.5 \log{T} (1 + \frac{1}{1 + \sqrt{m}})^2 \leq 12.5 \log{T}$, 
assuming $q_i(l_i) \leq \alpha q_j(l_j), \forall{i, j = 1, \dots, n}$, for some constant $\alpha > 1$, and $Q, q_i(l_i) \in Z, \forall{1 \leq i \leq n}$.

Note that this is an improvement over the algorithm in \cite{Armaselu-PETRA} when the aforementioned assumption holds.

\section{Conclusions and Future Work}
\label{s:conclusion}

We presented an APX for the Maximum-Profit Routing Problem with Variable Supply (MPRP-VS).
We also improve upon the results in \cite{Armaselu-PETRA} for the  Maximum-Profit Routing Problem (MPRP) under certain conditions.

We leave for future work proving lower bounds on the approximation ratio (assuming polynomial running time) for the problem,
as well as coming up with randomized algorithms that are either faster or have better approximation factor, with high probability. 

We also leave for future consideration finding approximation algorithms for the other versions of MPRP, listed below.

\textbf{MPRP-M}. More than one vehicle is allowed per site, and the quantity is fixed. 

\textbf{MPRP-MVS}. More than one vehicle is allowed per site, and the quantity varies in time. 

\section*{Acknowledgement}

The author would like to thank Dr. Ovidiu Daescu for helpful discussions.

\bigskip
\end{document}